# Pulsed Ultrasound Assisted Thermo-therapy for Subsurface Tumor Ablation: A numerical investigation


**Gajendra Singh**

PG Scholar

Department of Mechanical Engineering

National Institute of Technology Arunachal Pradesh, Arunachal Pradesh-791112, India

e-mail: gajendra.kashyap7@gmail.com

**Anup Paul[1]**

Assistant Professor

Department of Mechanical Engineering

National Institute of Technology Arunachal Pradesh, Arunachal Pradesh-791112, India

e-mail: catchapu@gmail.com

**Himanshu Shekhar**

Assistant Professor

Discipline of Electrical Engineering

Indian Institute of Technology Gandhinagar, Gujarat-382355, India

Email: himanshu.shekhar@iitgn.ac.in

**Abhijit Paul**

PhD Scholar

Department of Mechanical Engineering

National Institute of Technology Arunachal Pradesh, Arunachal Pradesh-791112, India

e-mail: abhijitpaul1501@gmail.com






## ABSTRACT

*High Intensity Focused Ultrasound (HIFU) is a promising therapy for thermal ablation and hyperthermia, characterised by it noninvasiveness, high penetration depth. Effective HIFU thermo-therapy requires the ability to accurately predict temperature elevation and corresponding thermal dose distribution in target tissues. We report a parametric numerical study of the thermal response and corresponding of thermal dose in a bio-tissue in response to ultrasound. We compared the predictions of tissue models with two, three and seven layers, to ultrasound induced heating at duty cycles ranging from 0.6 and 0.9. Further, two tumor sizes and transducer powers (10 W and 15 W) were considered. Inhomogeneous Helmholtz equation was coupled with Penne's bioheat equation to predict heating in response to pulsed ultrasound. Necrotic lesion size was calculated using the cumulative equivalent minute (CEM) thermal dose function. In-vitro experiments were performed with agar-based tissue phantoms as a preliminary validation of the numerical results. The simulations conducted with the seven layered model predicted up to 33.5% lower peak pressure amplitude than the three-layered model. As the ultrasound pulse width decreased with the equivalent sonication time fixed, the corresponding magnitude of the peak temperature and the rate of temperature rise decreased. Pulsed ultrasound resulted in increased the volume of necrotic lesions for equivalent time of sonication. The findings of this study highlight the dependence of HIFU-induced heating on target geometry and acoustic properties, and could help guide the choice of suitable ultrasound exposure parameters for further studies.*

**Keywords:** High intensity focused ultrasound, multi-layered tissue, duty cycle, thermal lesion.

---

[1]Corresponding Author





## 1. INTRODUCTION

Cancer is the second largest cause of death globally [1]. In recent years, treatments based on Radiotherapy [2], Laser Therapy [3,4], Cryoablation [5,6], microwave ablation [7,8], and focused ultrasound ablation [9,10] have been developed for cancer. Specifically, High Intensity Focused Ultrasound (HIFU) combines non-invasiveness, focusing ability, and high penetration depth [10–13]. Apart from cancer therapy, Focused Ultrasound has been used for applications such as drug delivery [14], blood brain barrier (BBB) opening [15], and sonothrombolysis [16]. Despite potential advantages, the widespread clinical adoption of HIFU necessitates optimisation of exposure parameters for rapid and homogeneous heating [17].

Focused ultrasound therapy is governed by acoustic and heat transfer physics. Numerical simulation is a powerful tool to predict the effectiveness of thermal therapy. Several reported studies have simulated the temperature variation due to HIFU using classical Fourier law based Penne's bioheat transfer equation [18–20]. Kumar et al.(2016) reported a Dual Phase Lag (DPL) model for simulating the response of multi-layer tissue composed of dermis, subcutaneous, fat, and muscle layers to HIFU [21]. Namakshenas et al. (2019) modelled thermal ablation of a thyroid tumor using HIFU [22]. Lin and colleagues (2013) published a study [23] based on the spheroidal beam equation and numerical alternate direction implicit method to assess the influence of ribs on therapeutic ultrasound propagation. Two-layer and multi-layered tissue phantoms have also been modelled with non-linear Westervelt equation to predict the





resulting pressure distribution [24]. Further, computer-based models have been developed to simulate the treatment of tumor using HIFU at moderate intensities [25].

Overall, previous studies have contributed to the modelling of linear and non-linear propagation effects on HIFU using different models such as ADI, SBE, Thermal wave model (TWMBT) and DPL model in a multi layered tissue. However, these studies have not explored the effect of HIFU with pulsed ultrasound exposure in multi-layered tissues, which is the focus of the present study. We report on the effect of ultrasound exposure parameters such as duty cycle and power on the temperature and thermal dose distribution. The simulations were carried out by considering five different computation domains. The present work also simulated the effect of tumor size on the time taken for treatment. To investigate the pressure distribution and temperature distribution in the tissue domain, acoustic physics was coupled with Fourier bioheat transfer equation. Experiments were conducted on heating of agar based tissue phantom using ultrasound transducer for prelimnary validation of the predicted results.

## 2. PROBLEM DEFINITION AND MATHEMATICAL MODELLING

### 2.1 Problem Definition and Physical Domain

The conversion of ultrasound into thermal energy depends upon parameters such as field intensity, frequency of excitation, exposure time and the properties of the tissue medium. An increase in temperature to more than 60°C for 1 s leads to protein denaturation which caused irreversible cellular damage in most tissues [10]. The configuration of the physical domain is shown schematically in Fig. 1 (a). The ultrasound





waves from the transducer was focused at the tumor. The empty space between the transducer surface and the tissue medium was modelled as water for adequate coupling [26]. To avoid the reflection of the acoustic waves at the boundary, a perfectly matched layer (PML) of 5 mm thickness was added in the domain [27].

The numerical simulations were performed using a 2D axisymmetric tissue model that is symmetric about z-axis to reduce the computational time. Five different tissue domains were considered viz., (a) two-layered water and tissue model (TLWT), (b) three-layered water and tumor-embedded tissue with small tumor (TLST), (c) three-layered water and tumor-embedded tissue with large tumor (TLLT), (d) seven-layered water and tumor-embedded tissue for small tumor (SLST) and (e) seven-layered water and tumor-embedded tissue for large tumor (SLLT) as shown in Fig. 1 ((b)-(d)). Elliptical tumors sizes of 7.5 x 3 mm were introduced in TLST and SLST domain and tumors of 15 x 6 mm were introduced in in TLLT and SLLT. We keep the distance to 20 mm [28]. The thickness of different layers, acoustic and thermal properties of the tissue are given in Table 1.

<Figure 1>

<Table: 1>





## 2.2 Governing Equations

### 2.2.1 Propagation of Ultrasound Wave

The inhomogeneous Helmholtz equation was solved to model the propagation of the ultrasound wave in the medium [29]. The generalised governing equation for the acoustic wave is described by eqs. (1), (2) [30,31] respectively as discussed below,

$$\nabla \cdot \left( -\frac{1}{\rho_c}(\nabla p_{tot} - q_d) \right) - \frac{K_{eq}^2}{\rho_c} p_{tot} = Q_m \tag{1}$$

$$K_{eq}^2 = \left( \frac{\omega}{c_c} \right)^2 + \left( \frac{m}{r} \right)^2 \tag{2}$$

where $\rho_c$ and $c_c$ represent complex density and complex speed of the sound respectively. The total acoustic pressure $(p_{tot})$ is the sum of the pressure solved for $p$, which is then equal to the scattered pressure $(\text{p} = \text{p}_s)$, and the background pressure $(p_b)$.

$$p_{tot} = p + p_b \tag{3}$$

The wave number used in the equations $(K_{eq})$ incorporates ordinary wave number $\left( \text{K} = \frac{\omega}{c_c} \right)$ as well as the out-of-plane wave number and circumferential wave number $(K_m = \frac{m}{r})$. Where $m, q_d$ and $Q_m$ are the circumferential mode number, dipole domain source and monopole domain source respectively. In the present case, $m, q_d$ and $Q_m$ have been assumed to be $0 \ (dimensionless), 0 \ (N/m^2)$ and $0 \ (s^{-2})$, respectively. Under these assumptions, the simplified pressure wave equation can be rewritten as [29];

$$\nabla \cdot \left( -\frac{1}{\rho_c} \nabla p_{tot} \right) - \frac{\omega^2}{\rho_c c_c^2} p_{tot} = 0 \tag{4}$$





$$\rho_c = \frac{\rho c^2}{c_c^2} \tag{5}$$

$$c_c = \frac{\omega}{(\omega/c) - i\alpha} \tag{6}$$

where $\omega$ is the angular frequency ($\omega = 2\pi f$) and $f$ is the frequency. $\rho$ , $c$ and $\alpha$ are the density, speed of sound and attenuation coefficient of the medium respectively.

### 2.2.2 Bio Heat Transfer for Temperature Distribution

To investigate the thermal distribution in terms of temperature response of tissue, Penne's bio heat transfer equation was solved. Traditional bioheat equation of Penne's [32] is given as

$$\rho_t C_t \frac{\partial T}{\partial t} = k_t \nabla^2 T - \omega_b \rho_b C_b (T - T_b) + Q_{ext} + Q_{met} \tag{7}$$

where $\rho$, $C$ and $k$ are the density, specific heat at constant pressure and thermal conductivity respectively. The subscript $t$ and $b$ stand for tissue medium and blood respectively. $\omega_b$ is the blood perfusion rate. $T$ and $T_b$ represents the tissue and arterial blood temperatures respectively.

The volumetric heat generation source ($Q_{ext}$) is proportional to the time average intensity magnitude ($I_{avg}$) and attenuation coefficient of tissue medium ($\alpha$), which is gives by equation (8) [33] as

$$Q_{ext} = 2\alpha I_{avg} \tag{8}$$

$$\alpha = \alpha_0 \left(\frac{f}{f_0}\right)^n \tag{9}$$

$$I_{avg} = \left(\frac{p^2}{\rho c}\right) = \left|Re\left(\frac{1}{2}pv\right)\right| = \frac{\rho c v^2}{2} = \frac{P}{A} \tag{10}$$

$$v = 2\pi f x_{max} \tag{11}$$





where the attenuation coefficient $(\alpha)$ depends upon the excitation frequency $(f)$ and follows the power law for the biological tissue and is expressed by equation (9) [34]. Where $\alpha_0$ is the attenuation coefficient at the reference frequency $f_0 = 1\,MHz$. The power law exponent $(n)$ is assumed to be $n \approx 1$ at lower frequency [35]. The intensity magnitude $(I_{avg})$ can be expressed in terms of either the pressure amplitude $(p)$ or the particle velocity amplitude $(v)$ as given in equation (10) [33,36]. The velocity may be expressed by equation (11) where $x_{max}$ is the maximum displacement of the particle [36]. $P$ and $A$ are the transducer power and transducer area respectively. In the equation (7), the term $Q_{met}$ represents the metabolic heat generation rate. The term $Q_{met}$ is set to be zero because the generated metabolic heat is very small as compared to the heat generated by the external source $(Q_{ext})$ [37].

<Figure 2>

<Table: 2>

## 2.3 Initial and Boundary Conditions

A concave transducer with a centre frequency of 1 MHz, aperture diameter of 70 mm and 65 cm focal length was modelled as the source of ultrasound. Two different powers 10 W and 15 W [21] were considered. As the power was increased 1.5-fold (from 10 W to 15 W), the exposure time for 15 W power was reduced by the same factor. Exposure time at different powers and different duty cycles is summarised in Table 2.

An inward normal displacement that was harmonically oscillated was applied to the transducer surface as a boundary condition. It is an external source term that





couples acoustics with the structure. Displacement was converted to the acoustic pressure and used as the input parameter for the transducer. Except this boundary, other wall boundaries were defined by sound hard boundary (wall) condition. At the wall boundary, the normal derivative of the pressure was zero. Since the computational domain was symmetric about the z-axis, therefore axial symmetry boundary condition was considered. Normal displacement, sound hard wall and axial symmetry boundary conditions are shown in the Fig. 2(a) with Blue, Red and dashed line respectively.

For the thermal analysis, initial temperature of the tissue medium was assumed to be 37°C. A constant temperature of T=37°C was used to define the external boundaries of the tissue medium as shown in Fig. 2(a) by green colour. Heat source for tissue medium based on the equation (8) was assigned to get the thermal response of tissue.

For pulsed heat transfer analysis, heat source term was coupled with the ultrasound duty cycle. The transducer was activated by four different duty cycles (0.6, 0.7, 0.8 and 0.9), as shown in Fig. 2(b). For these duty cycles, the waveform with angular frequency of $2\pi$ and amplitude ratio of 1 were considered. The heating time for different duty cycles and powers is given in Table 2.

**2.4 Calculation of Cumulative Equivalent Thermal Dose**

To investigate the effect of elevated temperature on irreversible cellular damage, an empirical relation developed by Separeto and Dewey [38] was used, which depends on time temperature history and is given as





$$TD\ (t) = t_{43} = \int_{t_0}^{t_f} R^{(T(t)-316.15)}\ dt \tag{12}$$

Where $t_{43}$ is the equivalent time at 43℃. R is the isodose constant which is assumed to be 0.5 for temperatures above and equal to 43℃ or 0.25 for temperatures below 43℃. For prediction of lesion size, the knowledge of the thermal dose is required. For the complete necrosis of biological tissue, the value of thermal dose is considered to be in the range of 25-240 min [38,39].

<Figure 3>

## 3. NUMERICAL METHODOLOGY

For simulating acoustic and heat transfer physics, meshing with 62,315 and 57,908 triangular elements, respectively, was found to be satisfactory. These mesh elements were selected by performing grid independency test as shown in Fig. 3 (a) and (b). Mesh convergence was performed by changing the element size while keeping other parameters constant. Since the grid must solve the acoustic wave, wavelength dependent mesh was used (where $\lambda$ is the wavelength of the sound wave ($\lambda = c\ /\ f$ and $f = 1\ MHz$ ). The variation of pressure along axial direction for two different element sizes and was found to be nearly identical (Fig. 3(c)).

The Helmholtz equation and Penne's equation for thermal response along with initial and boundary conditions were solved by finite element method using COMSOL Multiphysics (COMSOL Inc, Bengaluru, India) software. The model was discretised using element order of quartic Lagrange for acoustic physics and linear for temperature field. Frequency domain and time dependent studies were used. All the studies had a relative





tolerance of $10^{-4}$. Steps taken by the solver were ``strict" using implicit Backward Differentiation Formula (BDF) method with maximum time step of 0.01. Since thermal dose function is the time integral of temperature, therefore it was solved with time dependent procedure. Simulation was conducted sequentially. First, the acoustic pressure was solved for the complete domain and then heat transfer was solved by coupling the acoustic pressure. The simulation was executed by Dell Precision Tower 7810 (Intel Xeon E5 v4 Processor, 20 Core, 96GB DDR4 RAM).

## 4. EXPERIMENTAL METHODOLOGY

The complete experiment has been performed in three steps:

(a) Preparation of tissue phantom

(b) Measurement of thermal and acoustic properties of prepared phantom samples

(c) Ultrasound generation and temperature monitoring.

### 4.1 Preparation of equivalent tissue phantom

For the preparation of a tissue phantom, 2.6 gm [3] of agar powder (Sisco Research Laboratories, Mumbai, India) was mixed in 100 ml of deionised water (Type1 Ultrapure Water, Evoqua Water Technologies India Private Limited, Chennai, India). Then the mixture was heated on Induction heater (HD4938/01, Philips, Gujarat, India) and stirred until the milky white solution became transparent and boiling began. This process took about 10-15 minutes. When the solution became transparent, then it was poured in a





test section (acrylic basin) and allowed to rest for 12 hours at room temperature to ensure complete solidification.

## 4.2 Measurement of thermal and acoustic properties

The thermal properties of the tissue phantom were measured by thermal property analyser (TEMPO, Meter Group, Inc., USA). The speed of sound and attenuation coefficient of the phantom was measured using a through-transmission acoustic setup [40]. A 1-MHz transducer (20 mm aperture, 50 inch focus, Roop Telsonic Ultrasonix Limited, Mumbai, India) was driven by a frequency generator (FY6900-60M, FeelTech, Henan Province, China) at 24 V and a confocally aligned facing a 2.5-MHz transducer (20 mm aperture, 50 inch focus, Roop Telsonic Ultrasonix Limited) was used to receive the signals. The phantom was placed centrally in between the transducers. To measure the attenuation, transducer was driven with a 30-cycle sineburst at 1 MHz frequency using frequency generator with and without tissue phantom. The received output signal was digitised using a storage oscilloscope (SMO120E, Scientific Mes Technik Private Limited, MP, India) to 12 bits and transferred to a computer for calculating the attenuation coefficient. To measure the speed of sound, a pulse generator (RTUL MHF-400, Roop Telsonic Ultrasonix Limited, Mumbai India) was used to drive the transducer and the signal from the received transducer was detected and visualised using the digital storage oscilloscope. The dimensions of the phantom were measured using digital callipers, and the time-shift obtained after placing the phantom in the path was measured and





analysed in MATLAB to calculate the speed of sound. The thermal and acoustic property of the tissue phantom is listed in Table 3.

<Table 3>

<Figure 4>

**4.3 Experimental Setup and Temperature Monitoring**

The complete experimental setup as shown in Fig. 4 consists of Ultrasound therapy unit (LCS-128, Life Care Systems, Uttar Pradesh, India), an acrylic cylindrical test section containing agar tissue phantom, K type thermocouples for temperature measurement, a data logger (34980A, Keysight Technologies Private Limited, Bangalore, India) and a computer system for monitoring. The transducer was driven using 15 W power with 1 MHz of frequency. It was turned on for 215 s for heating and turned off subsequently for cooling for another 325 s. The ultrasound transducer was held in place by a custom positioning system. The diameter and height of acrylic test section were 10 cm and 15 cm respectively with provision to insert thermocouples radially. Ultrasound gel was used as a coupling agent was used between the ultrasound probe and tissue phantom. The K-types thermocouples were position at the focal area of the tissue phantom to monitor the temperature. Finally computer systems along with a data logger were used for continuous monitoring and recording of phantom temperature.

<Figure 5>





**5. RESULTS**

**5.1 Validation**

The numerical results were compared with known solutions available in the literature [41]. For validation of the computational domain, acoustic and thermophysical properties and the other operating parameters like frequency, focal length and aperture were kept identical to a previous study.[41]. Following the approach of Abdolhosseinzadeh et al., [41], the simulation was performed at two different normal displacements of 5 nm and 1 nm keeping the frequency at 1 MHz. The axial variation of absolute acoustic pressure (as shown in Fig. 5(a)) was in agreement with previous work. The present numerical results were also verified with the multi-layered bio-tissue model subjected to focused ultrasound in terms of thermal power dissipation and temperature distribution [37]. The comparison is shown in Fig. 5(b) and (c). Figure 5(b) shows the axial variation of thermal power dissipation and the maximum temperature distribution, which were in agreement with the study reported by Gupta et al. (2019), [37].

Furthermore, preliminary experimental validation of the model was performed in this study. the numerical results were compared with the (discussed in section 4.3) temperature distribution measured experimentally. The properties used in the simulation were measured experimentally and are listed in Table 3. During the experiment, the agar based tissue phantom was sonicated for 215 s and then allowed to cool for next 325 s at 1 MHz frequency and power of 15 W. The comparison of measured maximum temporal temperature rise with that of simulation has been shown in Fig. 5(d) with a maximum difference of 9%.





<Figure 6>

## 5.2 Spatial and temporal thermal analysis of the tissue

The numerical simulation was employed to determine the thermal response of tissue under ultrasound exposure. Thermal analysis was carried out based on classical Fourier based Penne's bioheat equation, coupled with inhomogeneous Helmholtz equation (discussed in section 2). The solution provided by the Helmholtz equation was used as an external heat source in the Penne's bioheat equation. The result from the inhomogeneous Helmholtz equation is shown in Fig. 6. Figure 6(a) shows the maximum pressure for different tissue domains at 10 W and 15 W. Out of all five tissue models, TLST showed the maximum pressure amplitude whereas the SLLT had the minimum pressure amplitude at same power level. Three layered model demonstrated 33.5% and 31.1% higher pressure than seven layered model for large tumor and small tumor, respectively at both power levels. Figure 6 (b) and (c) show the intensity magnitude along the z-direction for 10 W and 15 W exposures, respectively. The same trend was obtained except a change the magnitude, likely because the other parameters were kept constant. The peak value was shifted away from the source of excitation for the seven layered model but still within the focal region. As the deviations in the path increased, the focal point shifted further by up to 1.9 mm. Figure 6 (d), (e), (f) shows equivalent heat generation from induced pressure (equation (8)) generated in the tissue medium corresponding to the intensity rise at 15 W for TLWT, TLLT and SLLT





respectively. The heat generation was predicted to be higher for TLLT compared to other two models due to higher intensity and higher pressure rise as given by eqs. (8) and (10).

<Figure 7>

The thermal response of the different tissue models is shown in Fig. 7. The variation in temperature elevation at the focal point for different duty cycles and tissue models is shown in column 1. A significant temperature difference between two-layered, three-layered and seven-layered tissue was predicted in the pulsed and continuous sonication regimes at the same power level. All models with the smaller tumor predicted higher peak temperature than the larger tumor for the three-layered and seven-layered tissue models. The temperature rise was found to be reduced further with decrease in pulse width of ultrasound heating. The three layered model (TLST, TLLT) also predicted higher temperature as compared to seven layered model (SLST, SLLT (Fig. 6)). The temporal temperature distribution at continuous, 0.9 and 0.8 duty cycle for TLWT, TLLT and SLLT tissue model is shown in column 2. TLLT showed the maximum temperature rise and TLWT had the minimum temperature rise at same power. The predicted transient temperature distribution for different duty cycles for SLLT model at 10 W (Row 1) and 15 W (Row 2) is shown in column 3. As the duty cycle of pulse decreases, the rate of temperature rise as well as peak temperature is also decreases for equivalent time of heating (Table 2). A sharp increase in rate of temperature rise was predicted in case of ultrasound heating with higher power. A





higher peak temperature was observed at 15 W than 10 W in lower equivalent time of heating.

<Figure 8>

The contour plots (Fig. 8) of temperature in the r-z plane are shown for SLST at 10 W (column1), SLLT at 10 W (Column2) and SLLT at 15 W (Column3) at different duty cycles (Row 1 to 4). The heating time for different duty cycles and powers is given in Table 2. An increase in magnitude of peak temperature was noticed with increase in duty cycle. The maximum temperature for SLST and SLLT for 10 W was predicted to be 97.9°C and 88.5°C respectively for continuous heating of 8.4 s. Although the magnitude of peak temperature in case of larger tumor was lower than the smaller tumor, the entire tumor area received the appropriate thermal dose for treatment, i.e., temperature greater than 43°C [38]. Similar effects can be seen in Row 1 to Row 4 for continuous heating, 0.9 duty cycle, 0.8 duty cycle and 0.7 duty cycle exposure, respectively. When the power was increased to 15 W and time of heating was reduced to 5.6 s (continuous) for SLLT (column 3, Row 1), the peak temperature increased to 100 °C.

<Figure 9>

<Figure 10>

Figure 9 depicts the maximum temperature distribution along radial (at focal point, z=65 mm) and axial direction (r=0) for SLLT at 10 W and 15 W. The simulation





results showed a significant reduction in peak temperature at both powers with decreasing pulse width. Notably, the heat spread increased approximately 4-fold in axial direction (Fig. 9(b), (d)) than in the radial direction (Fig. 9(a), (c)). The elevation of temperature within the entire tumor was well above the temperature required for effective treatment for all duty cycles (Fig. 9(b)). However, an increase in the temperature (up to 4.23 and 2.15 degrees) was observed in dermis and bone layer due to the higher magnitude of attenuation coefficient (Fig. 6(f)) than other tissue layers. Figure 10 depicts the effect of pulse width while keeping the maximum temperature constant for SLLT at 10 W and 15 W. An increase in thermal spread was predicted with decreasing pulse width (Fig. 10(a)). Interestingly, this effect was found to be absent in the axial direction as shown in in Fig. 10(b).

<Figure 11>

<Table 4>

**5.3 Prediction of equivalent thermal dose**

The main goal of thermal ablation is to ensure the complete necrosis of tumor. The ablated area using the thermal dose function (section 2.4) is illustrated in Fig. 11 for SLST and SLLT at 10 W, and SLLT at 15 W for continuous and pulsed mode of sonication. The cumulative equivalent time and the percentage of necrotic tumor volume for different pulsed exposure modes and the two tissue model are listed in Table 4.  The lesion sizes obtained at 0.8 duty cycle and continuous mode for SLST were identical, and





a slight difference (4.66 %) difference was observed for SLLT as shown in Fig. 11((a) and

(b)). The lesion size increases with decrease in duty cycle as shown in Fig. 11((b) and (c)).

Also higher power, the thermal dose was reduced by 37.2%, 43.9% and 46.8% for

continuous, 0.8 duty cycle and 0.6 duty cycle mode of sonication, respectively for SLLT.

## 6. DISCUSSION

In this paper, temperature elevation and corresponding thermal dose distribution in a

subsurface tumor was investigated numerically in response to ultrasound exposure.

Several models (two-layered (TLWT), three-layered (TLST and TLLT) and seven-layered

(SLST and SLLT) were used to predict the effect of pulsed ultrasound induced heating at

different duty cycles, transducer power, and two tumor sizes. The three-layered model

predicted higher peak pressure amplitude than the seven-layered model (Figure 6). The

seven layered model predicted upto 33.5% lower acoustic pressure amplitude in the

focal region when compared to the three-layered and two-layered models. These trends

are consistent with a previously published study [42]. Shifts in focal points were also

observed in the predictions from different models (Figure 6 (b) and (c)). These

differences are due to the differences in attenuation coefficients and speed of sound of

different tissue layers.

The profile of peak compression and rarefactional intensity distribution is seen and

simultaneously the magnitude of the peak value increased upto the focal point and then

reduced (Figure 6). The heat generation in the water layer was insignificant relative to

the tumor. The tumor had a higher absorption coefficient and hence the heat





generation rate was greater. During early stage of heating, the rate of temperature rise was higher, but decreased subsequently. This decrease in the rate of temperature rise was more prominent in case of TLWT as compared to SLLT and TLLT and reached steady state more rapidly.

It was shown that an increase in thermal spread is expected with decrease in pulse width (Fig. 10(a)). Interestingly, this effect was to be absent in the axial direction as shown in in Fig. 10(b). Nonetheless, pulsed activation of ultrasound energy may be useful at lower powers for larger-sized tumors.

We found that the lesion size increases with decrease in duty cycle, particularly for the larger sized tumor (Fig. 11((b) and (c))). Therefore to obtain effective necrosis of larger tumor lower pulse width is recommended. Although this strategy increases the total treatment time, the peak temperature achieved is lower, which may improve the safety of therapy. Moreover, thermal dose can be reduced by increasing ultrasound power.

Based on the thermal analysis, a significant peak temperature difference was observed in case of SLLT model relative to the TLWT and TLLT models (Figure 7, column 1). The seven-layered model predicted higher peak temperature than TLWT, but lower peak temperature than TLLT irrespective of power and duty cycle.

A limitation of this study is that a Linear Model (Helmholtz equation) was used to simulate acoustic wave propagation and cavitation effects were ignored. However, as the focal pressure and intensities used in this study were low, nonlinear propagation and cavitation effects can be neglected without incurring substantial errors [20,25,43]. Additionally, preliminary in-vitro experiments were performed in this study with tissue





phantoms as a step towards the validation of the numerical results. However, more elaborate experimental studies and in vivo studies are needed in future to assess the full potential of pulsed HIFU therapy.

## 7. CONCLUSION

To ascertain the efficacy of focused ultrasound wave assisted thermo-therapy it is important to estimate accurate temperature elevation and corresponding thermal dose distribution in tumor embedded biological tissue. In this paper, a numerical study was reported to evaluate the effect of ultrasound exposure parameters such duty-cycle and power, and model parameters such as geometry and tumor size on ultrasound-mediated heating. Based on the thermal analysis the seven layered model predicted higher peak temperature when it compared with TLWT, but revealed lower peak temperature as compared to TLLT irrespective of power and duty cycle. The results revealed the tradeoffs between the time of treatment, maximum temperature achieved, and treatment zone size. Specifically, it was demonstrated that as the ultrasound pulse width decreases, the corresponding magnitude of the peak temperature as well as the rate of temperature rise is also decreases at equivalent sonication time. Further, this study underscores the significance of geometry on model predictions. Differences in model geometry resulted in corresponding differences in peak pressure amplitude, necrotic lesion size, and shifting of focal region. Specifically, the three-layered model predicted up to 33.5% higher peak pressure amplitude within the tumor than the seven-layered model. Taken together, these findings highlight the





need for further studies for optimization of treatment parameters in HIFU therapy to enhance the safety and efficacy of treatment.

**ACKNOWLEDGMENT**

AP would like to acknowledge TEQIP-III for financially supporting the computational work. HS was supported by Internal Project Grant from IIT Gandhinagar.






**REFERENCES**

[1]     Siegel, R. L., Miller, K. D., and Jemal, A., 2020, "Cancer Statistics, 2020," CA. Cancer J. Clin., **70**(1), pp. 7–30.

[2]     Cardinal, J., Klune, J. R., Chory, E., Jeyabalan, G., Kanzius, J. S., Nalesnik, M., and Geller, D. A., 2008, "Noninvasive Radiofrequency Ablation of Cancer Targeted by Gold Nanoparticles," Surgery, **144**(2), pp. 125–132.

[3]     Paul, A., Narasimhan, A., Kahlen, F. J., and Das, S. K., 2014, "Temperature Evolution in Tissues Embedded with Large Blood Vessels during Photo-Thermal Heating," J. Therm. Biol., **41**(1), pp. 77–87.

[4]     Paul, A., and Paul, A., 2020, "In-Vitro Thermal Assessment of Vascularized Tissue Phantom in Presence of Gold Nanorods During Photo-Thermal Therapy," J. Heat Transfer.

[5]     Zhao, F., and Chen, Z., 2011, "Three-Dimensional Numerical Study on Freezing Phase Change Heat Transfer in Biological Tissue Embedded with Two Cryoprobes," J. Therm. Sci. Eng. Appl., **3**(3), pp. 1–7.

[6]     Sukumar, S., and Kar, S. P., 2020, "A Combined Conduction–Radiation Model for Analyzing the Role of Radiation on Freezing of a Biological Tissue," J. Therm. Sci. Eng. Appl., **12**(1), pp. 1–11.

[7]     Simon, C. J., Dupuy, D. E., and Mayo-Smith, W. W., 2005, "Microwave Ablation: Principles and Applications," Radiographics, **25**(SPEC. ISS.).

[8]     Liu, J., Zhu, L., and Xu, L. X., 2000, "Studies on the Three-Dimensional Temperature Transients in the Canine Prostate during Transurethral Microwave







Thermal Therapy," J. Biomech. Eng., **122**(4), pp. 372–379.

[9]     Kennedy, J. E., 2005, "High-Intensity Focused Ultrasound in the Treatment of Solid Tumours," Nat. Rev. Cancer, **5**(4), pp. 321–327.

[10]    Zhou, Y.-F., 2011, "High Intensity Focused Ultrasound in Clinical Tumor Ablation," World J. Clin. Oncol., **2**(1), p. 8.

[11]    Beik, J., Abed, Z., Ghoreishi, F. S., Hosseini-Nami, S., Mehrzadi, S., Shakeri-Zadeh, A., and Kamrava, S. K., 2016, "Nanotechnology in Hyperthermia Cancer Therapy: From Fundamental Principles to Advanced Applications," J. Control. Release, **235**, pp. 205–221.

[12]    Bhowmik, A., Repaka, R., Mishra, S. C., and Mitra, K., 2016, "Thermal Assessment of Ablation Limit of Subsurface Tumor during Focused Ultrasound and Laser Heating," J. Therm. Sci. Eng. Appl., **8**(1), pp. 1–12.

[13]    Hsiao, Y. H., Kuo, S. J., Tsai, H. Der, Chou, M. C., and Yeh, G. P., 2016, "Clinical Application of High-Intensity Focused Ultrasound in Cancer Therapy," J. Cancer, **7**(3), pp. 225–231.

[14]    Tachibana, K., and Tachibana, S., 2001, "The Use of Ultrasound for Drug Delivery," Echocardiography, **18**(4), pp. 323–328.

[15]    Treat, L. H., McDannold, N., Zhang, Y., Vykhodtseva, N., and Hynynen, K., 2012, "Improved Anti-Tumor Effect of Liposomal Doxorubicin After Targeted Blood-Brain Barrier Disruption by MRI-Guided Focused Ultrasound in Rat Glioma," Ultrasound Med. Biol., **38**(10), pp. 1716–1725.

[16]    Shekhar, H., Kleven, R. T., Peng, T., Palaniappan, A., Karani, K. B., Huang, S.,






McPherson, D. D., and Holland, C. K., 2019, "In Vitro Characterization of Sonothrombolysis and Echocontrast Agents to Treat Ischemic Stroke," Sci. Rep., **9**(1), pp. 1–13.

[17] Zhu, L., Altman, M. B., Laszlo, A., Straube, W., Zoberi, I., Hallahan, D. E., and Chen, H., 2019, "Ultrasound Hyperthermia Technology for Radiosensitization," Ultrasound Med. Biol., **45**(5), pp. 1025–1043.

[18] He, Z. Z., Xue, X., and Liu, J., 2013, "An Effective Finite Difference Method for Simulation of Bioheat Transfer in Irregular Tissues," J. Heat Transfer, **135**(7), pp. 1–9.

[19] Jafarian Dehkordi, F., Shakeri-Zadeh, A., Khoei, S., Ghadiri, H., and Shiran, M.-B., 2013, "Thermal Distribution of Ultrasound Waves in Prostate Tumor: Comparison of Computational Modeling with In Vivo Experiments," ISRN Biomath., **2013**, pp. 1–4.

[20] Sheu, T. W. H., Solovchuk, M. A., Chen, A. W. J., and Thiriet, M., 2011, "On an Acoustics-Thermal-Fluid Coupling Model for the Prediction of Temperature Elevation in Liver Tumor," Int. J. Heat Mass Transf., **54**(17–18), pp. 4117–4126.

[21] Kumar, D., and Rai, K. N., 2016, "A Study on Thermal Damage during Hyperthermia Treatment Based on DPL Model for Multilayer Tissues Using Finite Element Legendre Wavelet Galerkin Approach," J. Therm. Biol., **62**, pp. 170–180.

[22] Namakshenas, P., and Mojra, A., 2019, "Numerical Study of Non-Fourier Thermal Ablation of Benign Thyroid Tumor by Focused Ultrasound (FU)," Biocybern. Biomed. Eng., **39**(3), pp. 571–585.






[23]    Lin, J., Liu, X., and Gong, X., 2013, "Computational Study on the Propagation of Strongly Focused Nonlinear Ultrasound in Tissue with Rib-like Structures," **134**(2).

[24]    Gupta, P., and Srivastava, A., 2018, "Numerical Analysis of Thermal Response of Tissues Subjected to High Intensity Focused Ultrasound," Int. J. Hyperth., **35**(1), pp. 419–434.

[25]    Hariharan, P., Myers, M. R., and Banerjee, R. K., 2007, "HIFU Procedures at Moderate Intensities - Effect of Large Blood Vessels," Phys. Med. Biol., **52**(12), pp. 3493–3513.

[26]    Casarotto, R. A., Adamowski, J. C., Fallopa, F., and Bacanelli, F., 2004, "Coupling Agents in Therapeutic Ultrasound: Acoustic and Thermal Behavior," Arch. Phys. Med. Rehabil., **85**(1), pp. 162–165.

[27]    Assi, H., and Cobbold, R. S., 2017, "Compact Second-Order Time-Domain Perfectly Matched Layer Formulation for Elastic Wave Propagation in Two Dimensions," Math. Mech. Solids, **22**(1), pp. 20–37.

[28]    Lang, B. H., and Wu, A. L. H., 2018, "The Efficacy and Safety of High-Intensity Focused Ultrasound Ablation of Benign Thyroid Nodules," Ultrasonography, **37**(2), pp. 89–97.

[29]    Huttunen, T., Malinen, M., Kaipio, J. P., White, P. J., and Hynynen, K., 2005, "A Full-Wave Helmholtz Model for Continuous-Wave Ultrasound Transmission," IEEE Trans. Ultrason. Ferroelectr. Freq. Control, **52**(3), pp. 397–409.

[30]    Comsol, 2010, "Acoustics Module," Interfaces (Providence)., pp. 1–218.

[31]    COMSOL Multiphysics, 2015, "Heat Transfer Module," Manual, pp. 1–222.







[32]    PENNES, H. H., 1948, "Analysis of Tissue and Arterial Blood Temperatures in the Resting Human Forearm," J. Appl. Physiol., **1**(2), pp. 93–122.

[33]    Nyborg, W. L., 1981, "Heat Generation by Ultrasound in a Relaxing Medium," J. Acoust. Soc. Am., **70**(2), pp. 310–312.

[34]    Szabo, T. L., 1995, "Causal Theories and Data for Acoustic Attenuation Obeying a Frequency Power Law," J. Acoust. Soc. Am., **97**(1), pp. 14–24.

[35]    Damianou, C. A., Hynynen, K., and Fan, X., 1995, "Evaluation of Accuracy of a Theoretical Model for Predicting the Necrosed Tissue Volume during Focused Ultrasound Surgery," IEEE Trans. Ultrason. Ferroelectr. Freq. Control, **42**(2), pp. 182–187.

[36]    Mamvura, T. A., Iyuke, S. E., and Paterson, A. E., 2018, "Energy Changes during Use of High-Power Ultrasound on Food Grade Surfaces," South African J. Chem. Eng., **25**, pp. 62–73.

[37]    Gupta, P., and Srivastava, A., 2019, "Non-Fourier Transient Thermal Analysis of Biological Tissue Phantoms Subjected to High Intensity Focused Ultrasound," Int. J. Heat Mass Transf., **136**, pp. 1052–1063.

[38]    Sapareto, S. A., and Dewey, W. C., 1984, "Thermal Dose Determination in Cancer Therapy," Int. J. Radiat. Oncol. Biol. Phys., **10**(6), pp. 787–800.

[39]    Damianou, C. A., Hynynen, K., and Fan, X., 1995, "Evaluation of Accuracy of a Theoretical Model for Predicting the Necrosed Tissue Volume during Focused Ultrasound Surgery," IEEE Trans. Ultrason. Ferroelectr. Freq. Control, **42**(2), pp. 182–187.







[40]   Shekhar, H., Smith, N. J., Raymond, J. L., and Holland, C. K., 2018, "Effect of Temperature on the Size Distribution, Shell Properties, and Stability of Definity®," Ultrasound Med. Biol., **44**(2), pp. 434–446.

[41]   Abdolhosseinzadeh, A., Mojra, A., and Ashrafizadeh, A., 2019, "A Numerical Study on Thermal Ablation of Brain Tumor with Intraoperative Focused Ultrasound," J. Therm. Biol., **83**(January), pp. 119–133.

[42]   Fan, T. B., Liu, Z. B., Zhang, Z., Zhang, D., and Gong, X. F., 2009, "Modeling of Nonlinear Propagation in Multi-Layer Biological Tissues for Strong Focused Ultrasound," Chinese Phys. Lett., **26**(8), pp. 6–9.

[43]   Hallaj, I. M., and Cleveland, R. O., 1999, "FDTD Simulation of Finite-Amplitude Pressure and Temperature Fields for Biomedical Ultrasound," J. Acoust. Soc. Am., **105**(5), pp. L7–L12.

[44]   McIntosh, R. L., and Anderson, V., 2010, "A Comprehensive Tissue Properties Database Provided for the Thermal Assessment of a Human at Rest," Biophys. Rev. Lett., **5**(3), pp. 129–151.

[45]   Kyriakou, A., Neufeld, E., Werner, B., Székely, G., and Kuster, N., 2015, "Full-Wave Acoustic and Thermal Modeling of Transcranial Ultrasound Propagation and Investigation of Skull-Induced Aberration Correction Techniques: A Feasibility Study," J. Ther. Ultrasound, **3**(1), pp. 1–18.

[46]   Chivers, R. C., and Parry, R. J., 1978, "Ultrasonic Velocity and Attenuation in Mammalian Tissues," J. Acoust. Soc. Am., **63**(3), pp. 940–953.

[47]   Duck, F. A., 1990, "Acoustic Properties of Tissue at Ultrasonic Frequencies," Phys.







Prop. Tissues, pp. 73–135.

[48]  El-Brawany, M. A., Nassiri, D. K., Terhaar, G., Shaw, A., Rivens, I., and Lozhken, K., 2009, "Measurement of Thermal and Ultrasonic Properties of Some Biological Tissues," J. Med. Eng. Technol., **33**(3), pp. 249–256.

[49]  Johnston, R. L., Dunn, F., and Goss, S. A., 1980, "Compilation of Empirical Ultrasonic Properties of Mammalian Tissues. Ll," J. Acoust. Soc. Am., **68**(1), pp. 93–108.






**Figure Captions List**

Fig. 1          Schematic of physical domain (a) multi-layered configuration; (b) two-layered water and tissue model (TLWT); (c) three-layered water and tumor-embedded tissue with small tumor (TLST) or large tumor (TLLT); (d) seven-layered water and tumor embedded tissue including dermis, subcutaneous, fat, muscle and bone for small tumor (SLST) or large tumor (SLLT).

Fig. 2          (a) Schematic of physical domain with boundary conditions (b) representation of pulse wave with duty cycle of 0.9, 0.8, 0.7 and 0.6.

Fig. 3          Grid Independency test in terms of (a) acoustic pressure; (b) maximum temperature; (c) pressure distribution along axial direction for two different mesh conditions.

Fig. 4          Illustration of experimental setup consists of ultrasound therapy unit, transducer probe, tissue phantom, thermocouple, data acquisition unit and computer.

Fig. 5          Comparison of the results of present numerical simulation (a) with Abdolhosseinzadeh et. al., [41] for absolute Pressure along axial direction; (b) with Gupta et. al., [37] for thermal power dissipation along z-axis ; (c) with Gupta et. al., [37] for transient temperature distribution ; and (d) with experimental observations (at 1 MHz, 15 W).

Fig. 6          (a) Maximum pressure for different tissue domains at 10 W and 15 W;





intensity magnitude along axial direction for TLWT, TLLT and SLLT tissue media at (b) 10 W and (c) 15 W; heat generation and average intensity along z-axis at 15 W for (d) TLWT, (e) TLLT (e) and (f) SLLT.

Fig. 7       (Column 1) Temperature elevation at different duty cycles for different physical models at 10 W (Row1) and 15 W (Row2); (Column 2) temporal temperature distribution at continuous mode, 0.9duty cycle, 0.8 duty cycle for TLWT, TLLT and SLLT at 10 W (Row1) and 15 W (Row2); (Column 3) temporal temperature distribution for SLLT at different pulse mode at 10 W (Row1) 15 W (Row2).

Fig. 8       Tissue thermal history on r-z plane for SLST at 10 W (Column1); SLLT at 10 W (Column 2) and 15 W (Column 3); with continuous (Row1); 0.9duty cycle (Row2); 0.8duty cycle (Row3); 0.7duty cycle (Row4) mode of heating.

Fig. 9       Temperature distribution at equivalent heating time for SLLT along radial direction (a, c) and axial direction (b, d) at 10 W (a, c) and 15 W (b, d).

Fig. 10      Temperature distribution at fixed maximum temperature for SLLT along radial direction (a, c) and axial direction (b, d) at 10 W (a, b) and 15 W (c, d).

Fig. 11      Thermal dose distribution on r-z plane for (a) SLST, (b, c) SLLT at 10 W (a, b) and 15 W (c) with different modes of sonication.





**Table Caption List**

Table 1     Thickness of tissue layers and their corresponding acoustic and thermal properties [12,21,22,37,44–49]

Table 2     Ultrasound exposure time at 10 W and 15 W.

Table 3     Measured acoustic and thermal properties of equivalent tissue phantom.

Table 4     Thermal dose and percentage of necrotic tumor for different tissue model, power and sonication conditions.





**Table 1**

| Layer/property | Water | Dermis | Subcutaneous | Fat | Muscle | Bone | Tumor |
|---|---|---|---|---|---|---|---|
| Thickness (mm) | -- | 4 | 8 | 13 | 25 | 20 | -- |
| Density, $\rho$ (kg/m$^3$) | 1000 | 1200 | 1000 | 911 | 1090 | 1908 | 1056 |
| Speed of Sound, c (m/s) | 1483 | 1624 | 1477 | 1440 | 1588 | 3515 | 1500 |
| 1Attenuation coefficient, $\alpha$ (Np/m) | 0.025 | 21.158 | 7 | 4.36 | 7.2 | 54.553 | 17.82 |
| Thermal conductivity, k (W/m.K) | -- | 0.4 | 0.21 | 0.21 | 0.5 | 0.32 | 0.53 |
| Specific heat, C (J/kg.K) | -- | 3500 | 2348 | 2348 | 3421.2 | 1313 | 3852 |
| Blood perfusion, $\omega_b$(1/s) | - | 0.0002 | 0.0001 | 0.0002 | 0.0005 | - | 0.0063 |





**Table 2**

| Power | Continuous (s) | 0.9 Duty Cycle (s) | 0.8 Duty Cycle (s) | 0.7 Duty Cycle (s) | 0.6 Duty Cycle (s) |
|-------|----------------|--------------------|--------------------|--------------------|--------------------|
| 10W | 8.4 | 9.3 | 10.5 | 12 | 14 |
| 15W | 5.6 | 6.2 | 7 | 8 | 9.4 |





**Table 3**

| Property | Phantom |
|---|---|
| Density, $\rho$ (kg/m$^3$) | 1050±0.016 |
| Speed of Sound, c (m/s) | 1500±0.02 |
| Attenuation coefficient, α (Np/m) | 0.7±0.017 |
| Thermal conductivity, k (W/m.K) | 0.66±0.004 |
| Specific heat, $c_p$ (J/kg.K) | 4219±40.32 |





**Table 4**

| Tissue Model | Mode of sonication | Thermal dose (s) | Percentage of the necrotic tumor (%) |
|---|---|---|---|
| SLST at 10W | Continuous | 8.2 | 45.05 |
| | 0.8 duty cycle | 8.6 | 45.85 |
| | **0.6 duty cycle** | **28.4** | **55.97** |
| SLLT at 10W | Continuous | 43 | 53.44 |
| | 0.8 duty cycle | 66 | 58.10 |
| | **0.6 duty cycle** | **237** | **69.90** |
| SLLT at 15W | Continuous | 27 | 51.37 |
| | 0.8 duty cycle | 37 | 54.55 |
| | **0.6 duty cycle** | **126** | **65.18** |





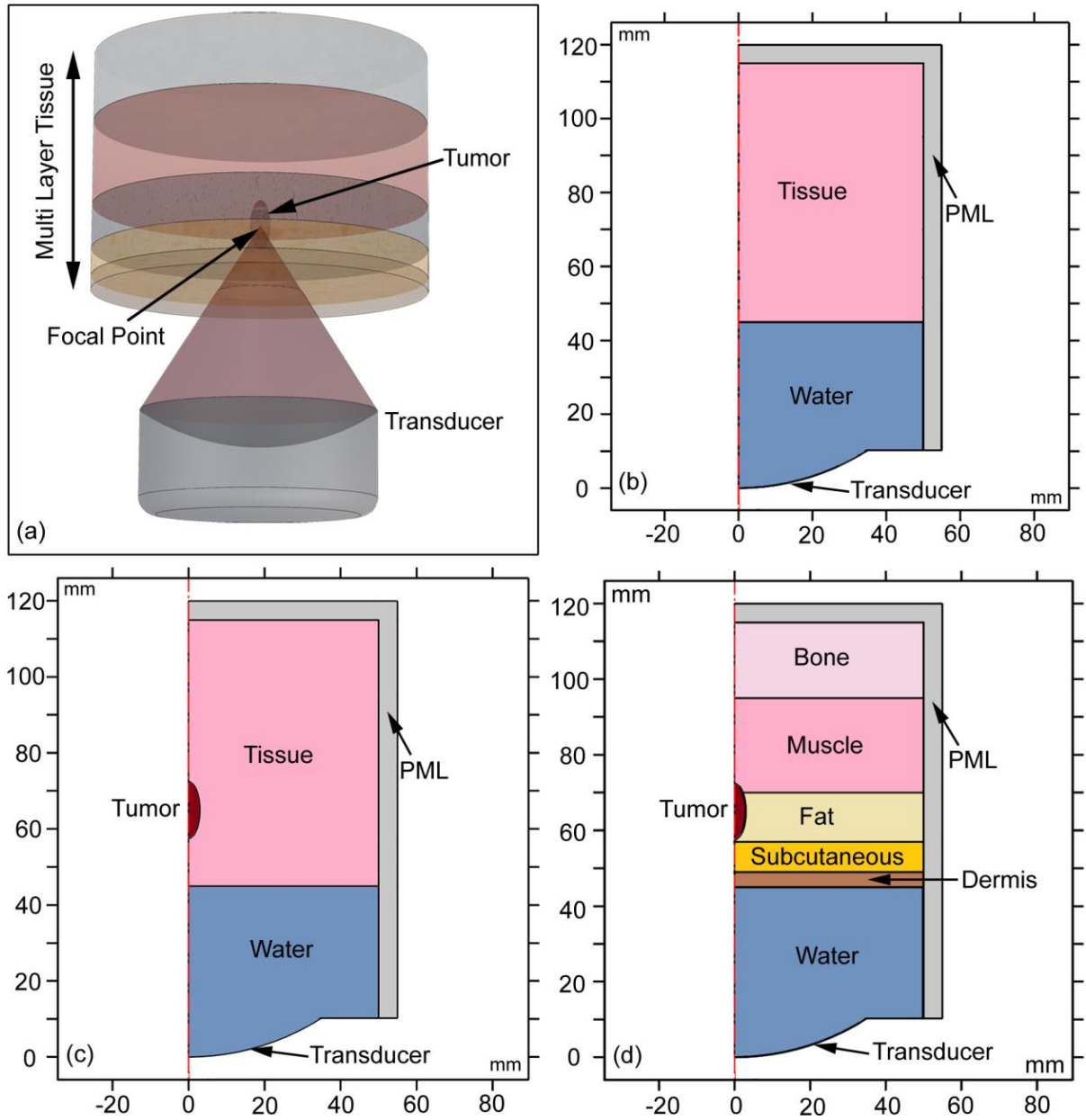

**Fig. 1**





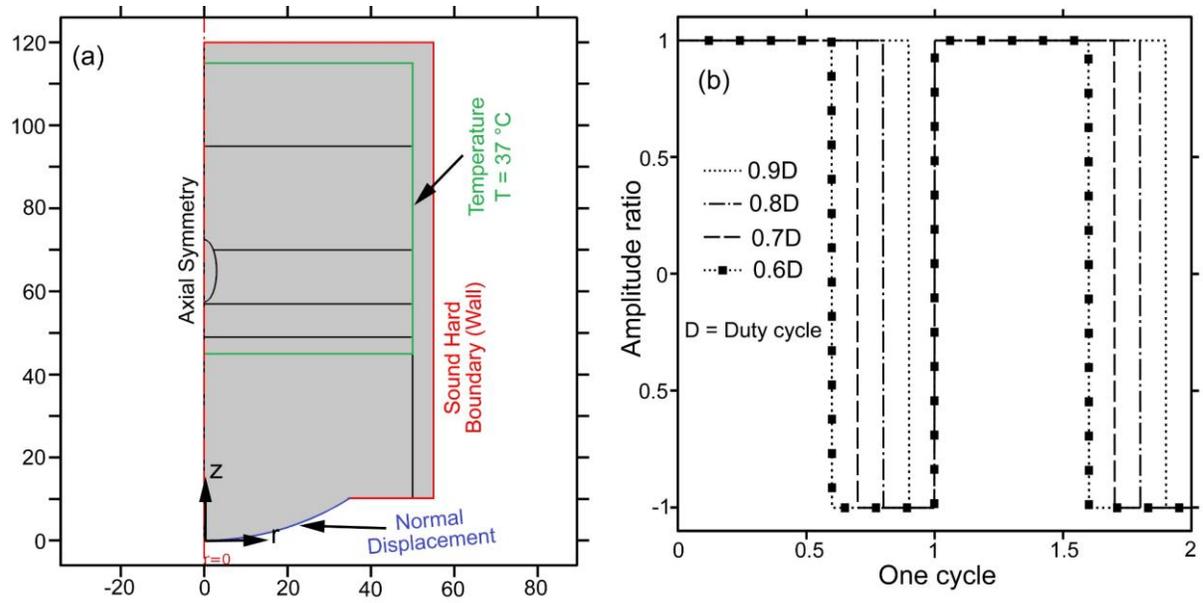

**Fig. 2**





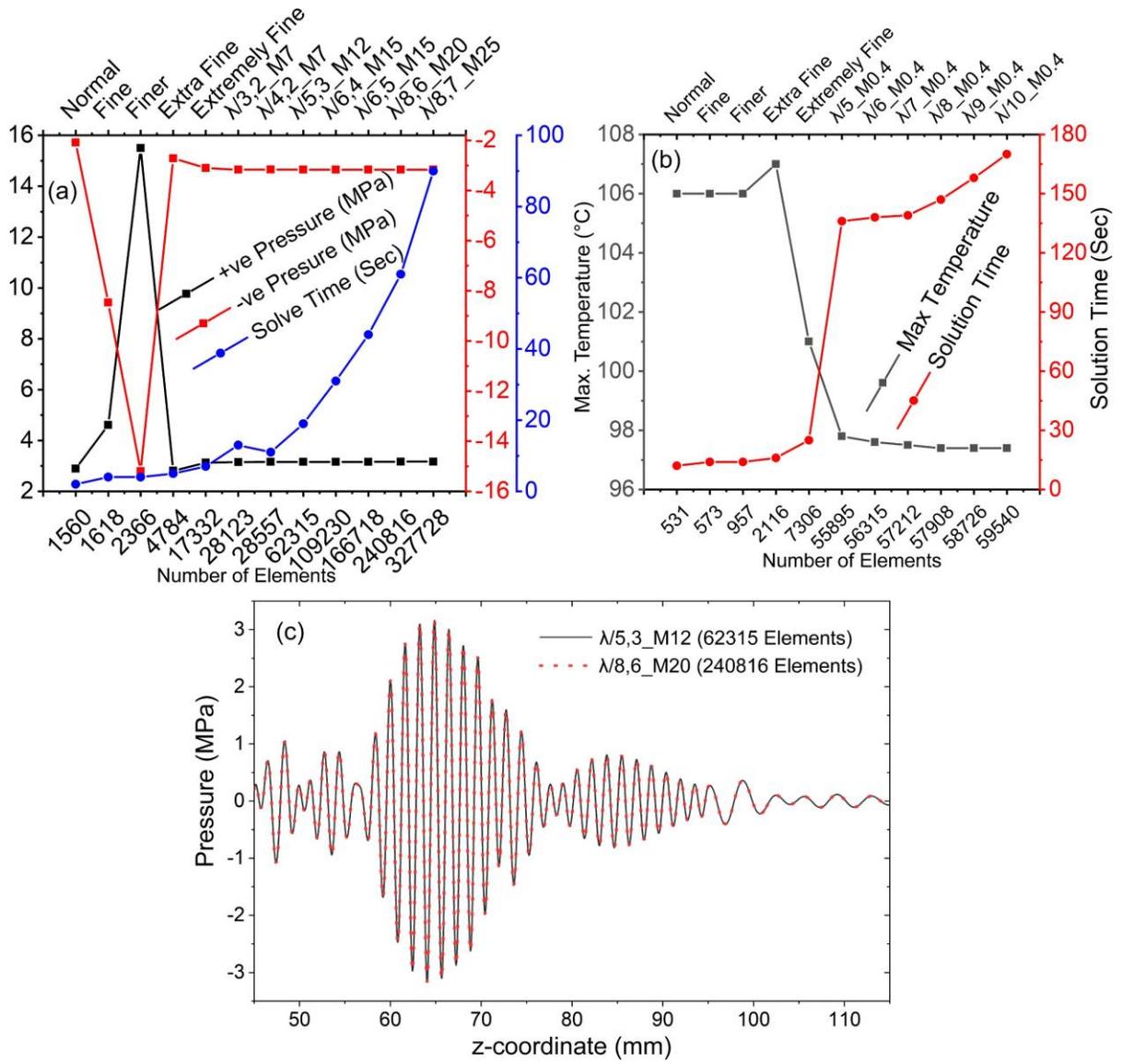

**Fig. 3**





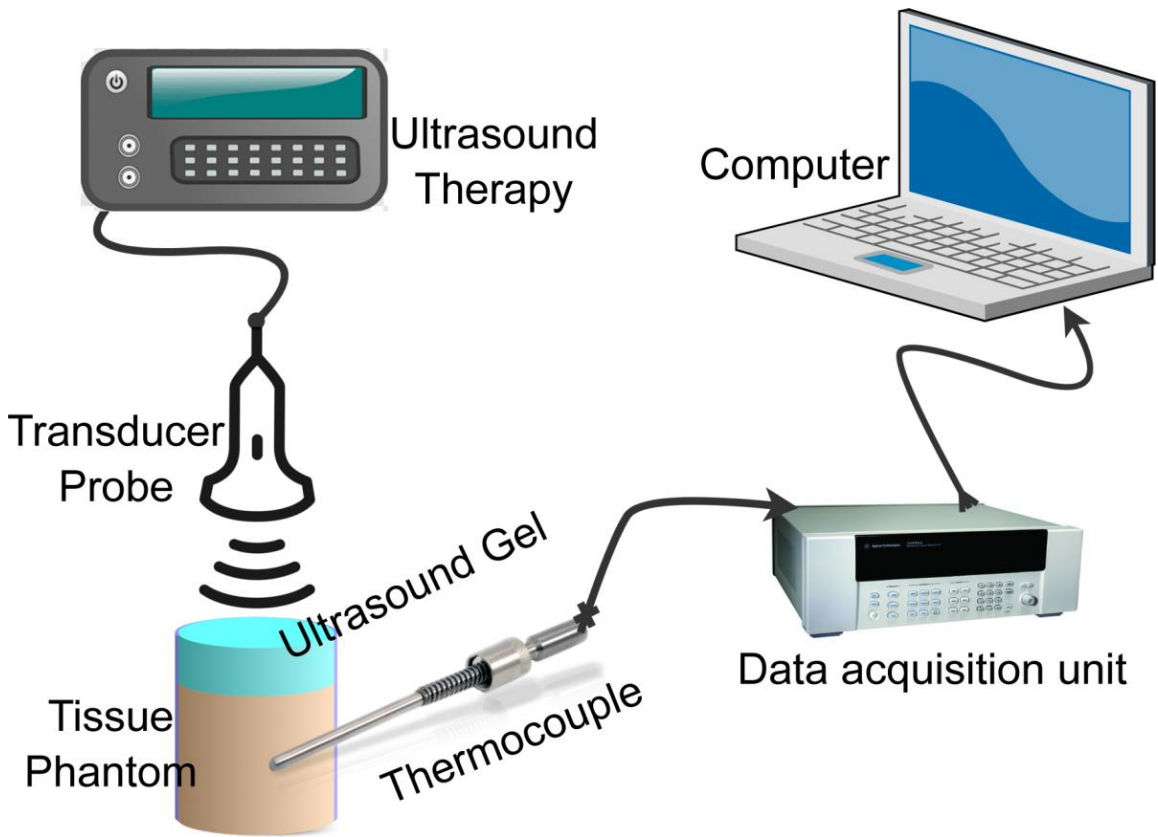

**Fig. 4**





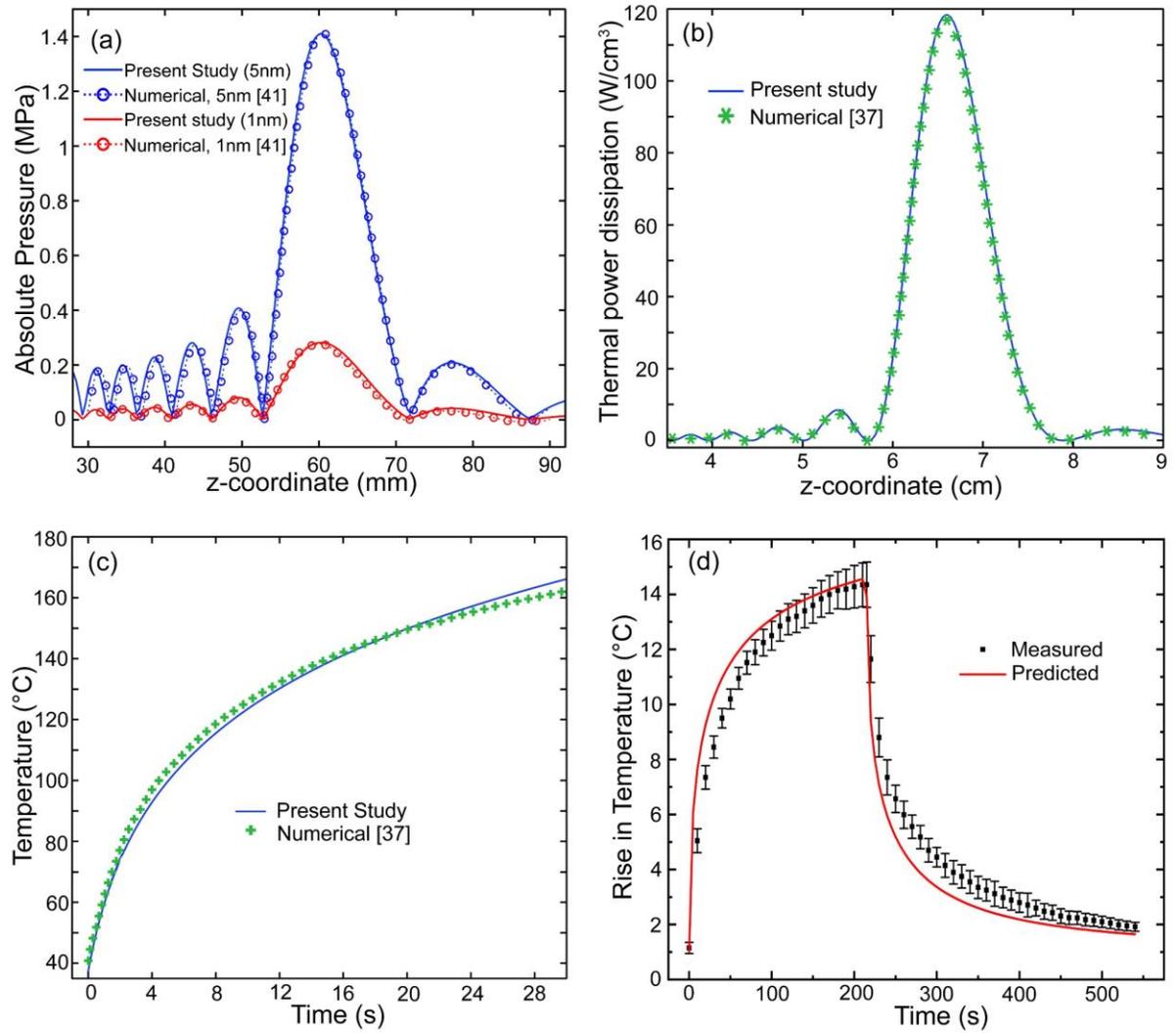

**Fig. 5**





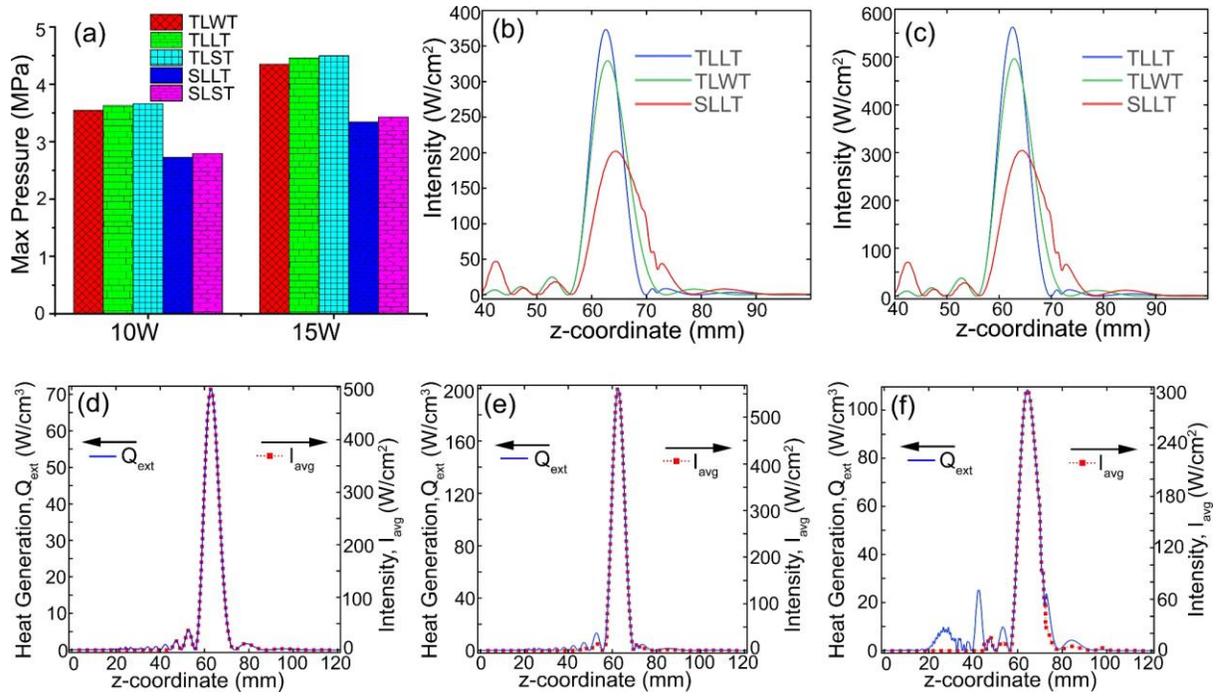

**Fig. 6**





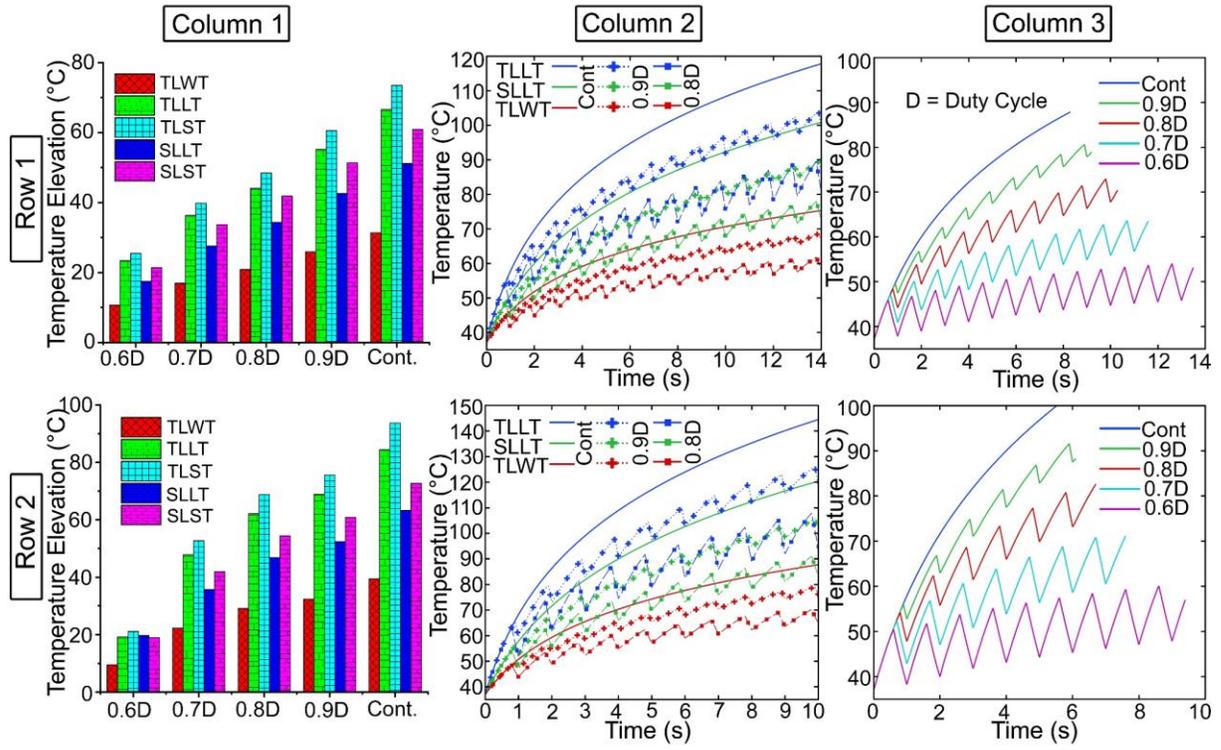

**Fig. 7**





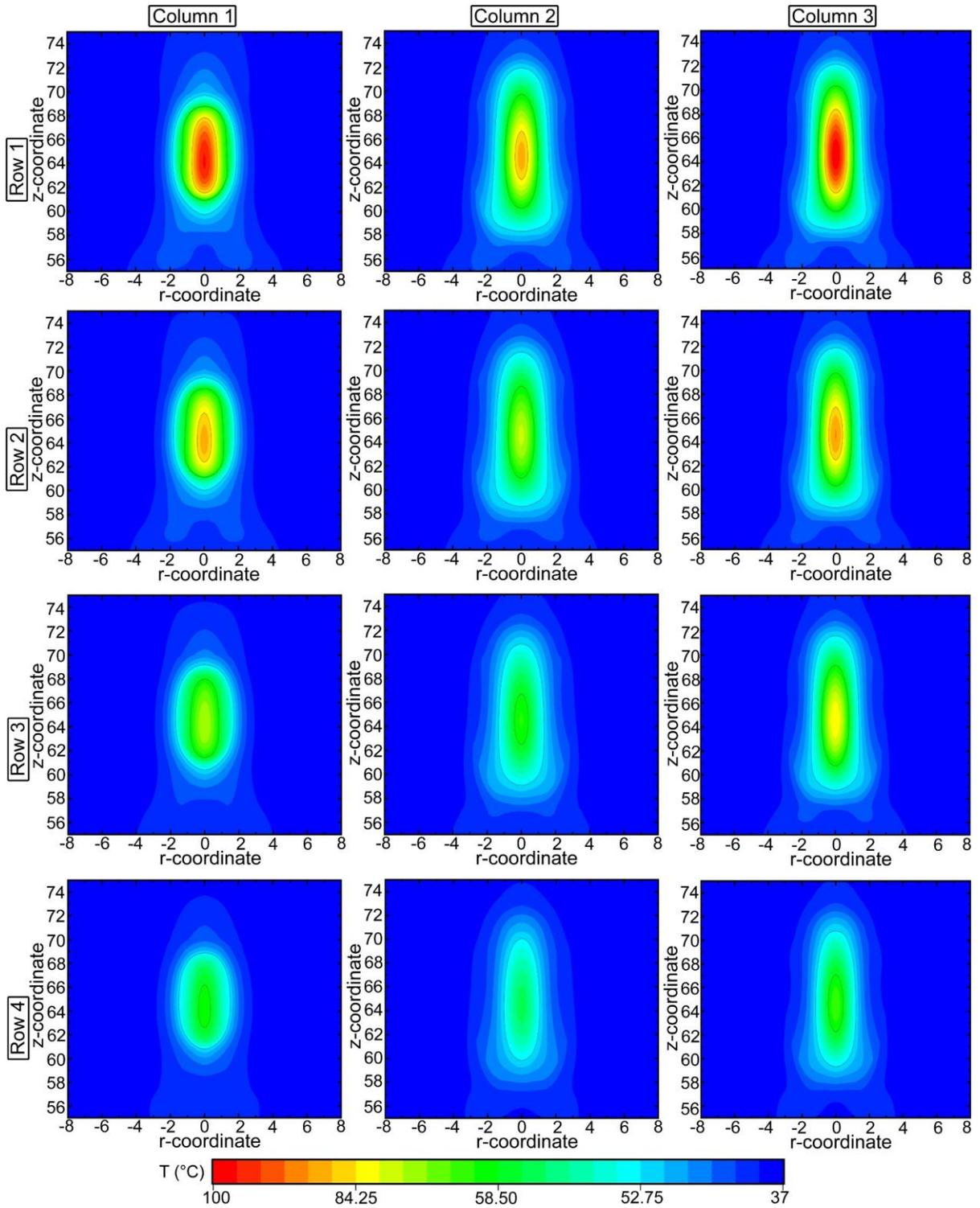

**Fig. 8**





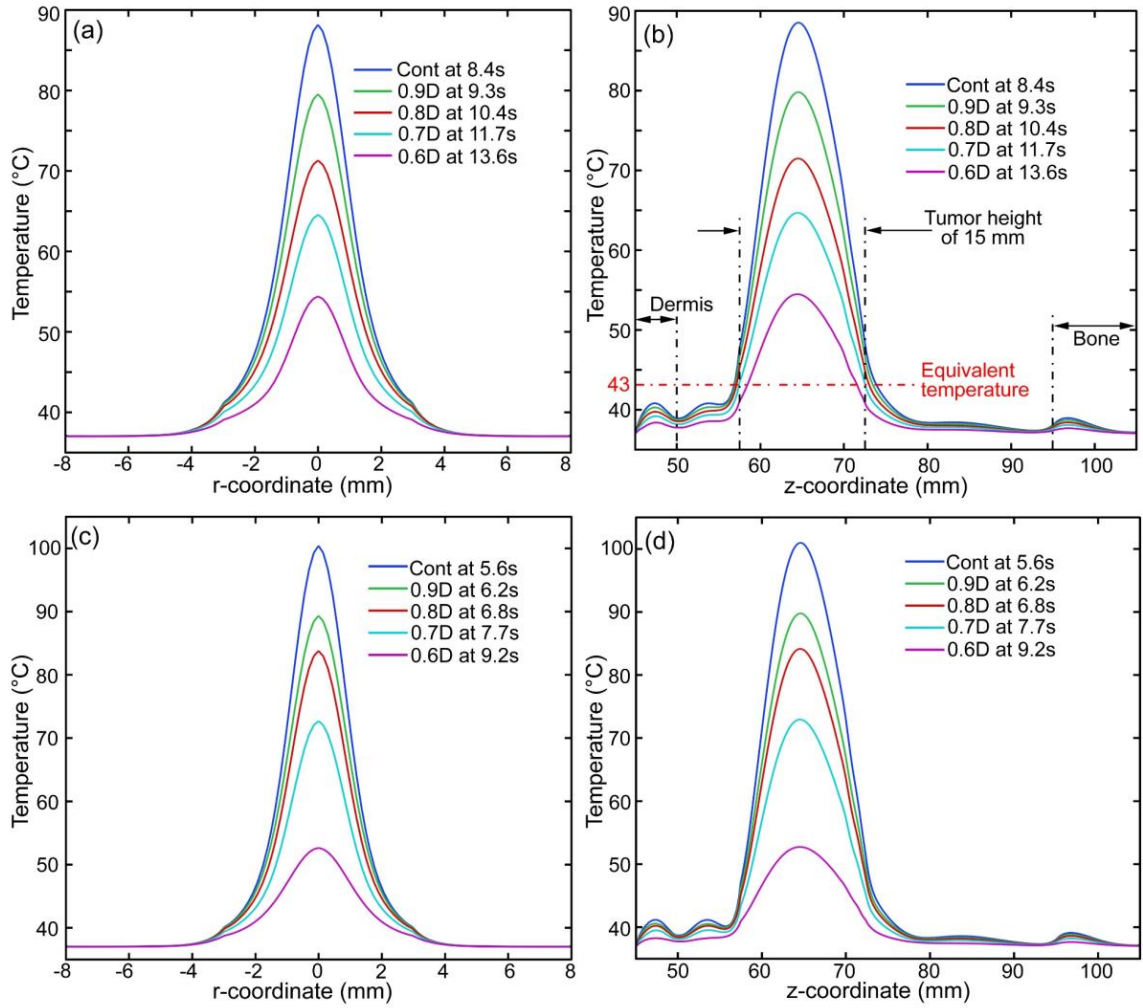

**Fig. 9**





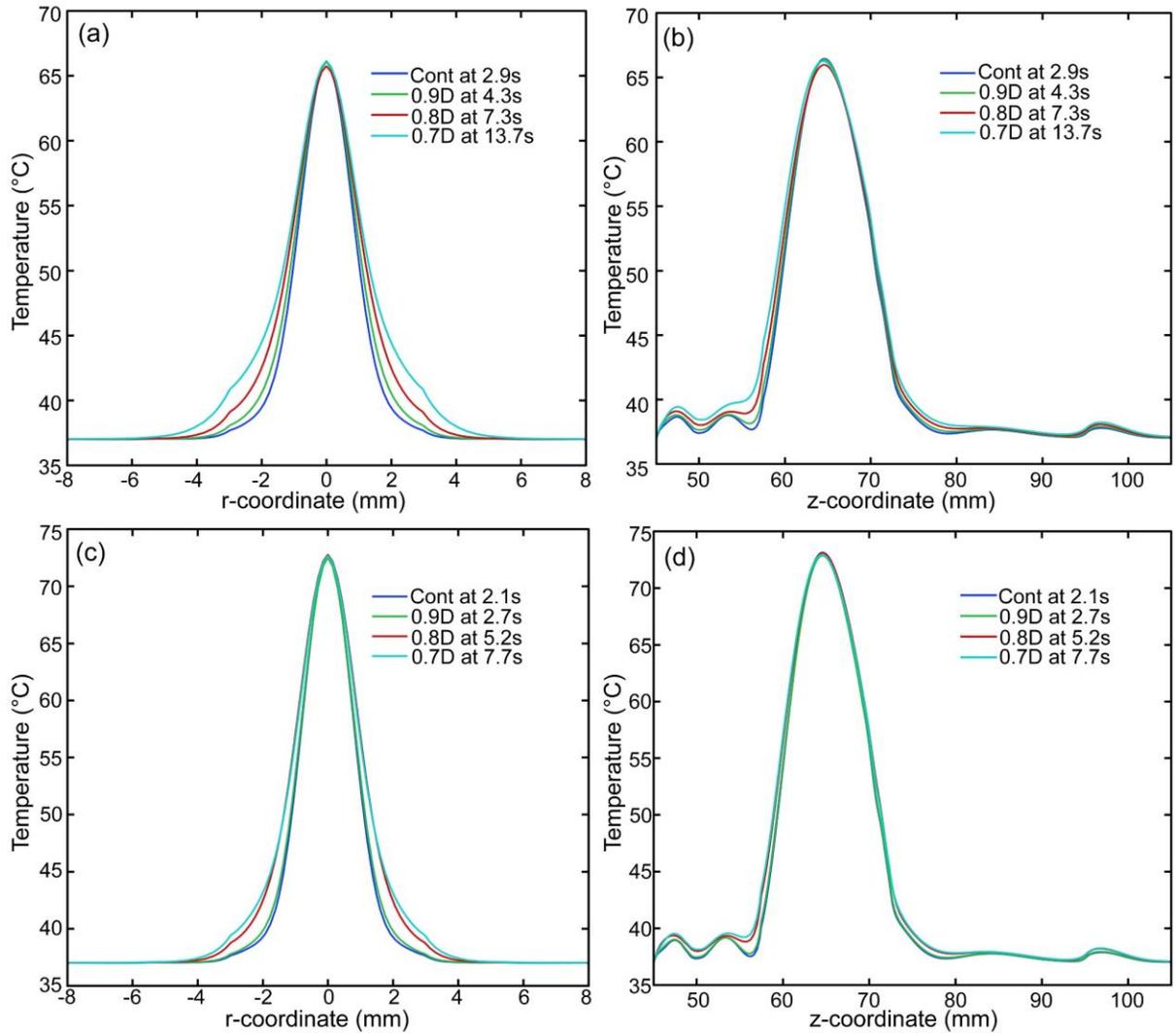

**Fig. 10**





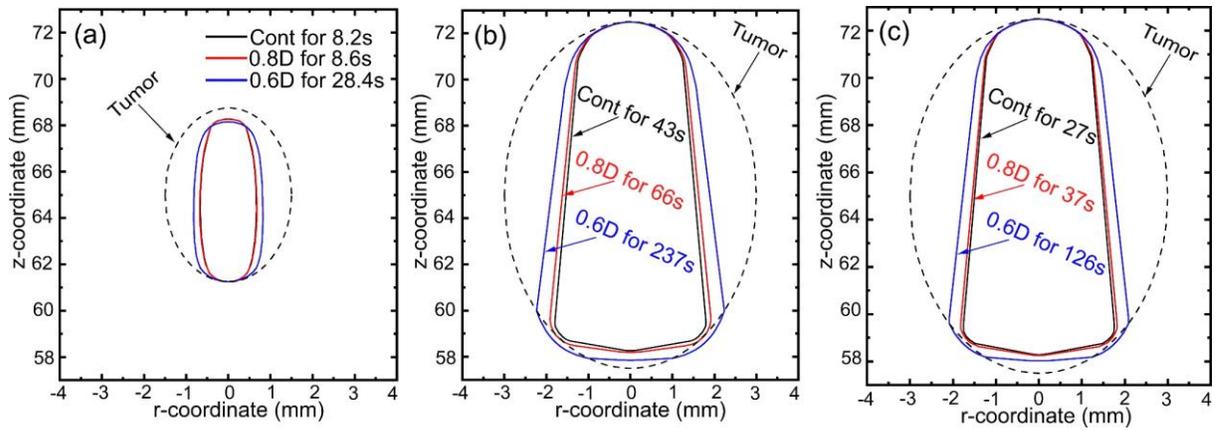

**Fig. 11**